\documentclass[iicol,pdflatex,sn-basic]{sn-jnl}

\usepackage{graphicx}%
\usepackage{multirow}%
\usepackage{amsmath,amssymb,amsfonts}%
\usepackage{amsthm}%
\usepackage{mathrsfs}%
\usepackage[title]{appendix}%
\usepackage[table,xcdraw]{xcolor}%
\usepackage{textcomp}%
\usepackage{manyfoot}%
\usepackage{booktabs}%
\usepackage{algorithm}%
\usepackage{algorithmicx}%
\usepackage{algpseudocode}%
\usepackage{listings}%
\usepackage{enumerate}
\usepackage{enumitem}
\usepackage{mathtools}
\usepackage{tabularx} 
\usepackage{mathrsfs} 
\usepackage{lmodern}

\usepackage[utf8]{inputenc}

\usepackage{cleveref}
\crefname{figure}{Fig.}{Figs.}
\crefname{section}{Sec.}{Secs.}
\crefname{equation}{Eq.}{Eqs.}
\crefname{table}{Table}{tables}
\Crefname{figure}{Fig.}{Figs.}
\Crefname{section}{Sec.}{Secs.}
\Crefname{equation}{Eq.}{Eqs.}
\Crefname{table}{Table}{Tables}

\newcommand{\ignore}[1]{}

\newcommand{\bol}{\boldsymbol}

\newcommand{\elf}{\mathbf E}

\newcommand{\rone}{\hat \rho_{1}}
\newcommand{\rtwo}{\hat \rho_{2}}

\begin{document}

\title{Differentiating through binarized topology changes: Second-order subpixel-smoothed projection}

\author*[1]{\fnm{Giuseppe} \sur{Romano}}\email{romanog@mit.edu}

\author[2]{\fnm{Rodrigo} \sur{Arrieta}}

\author[2]{\fnm{Steven G.} \sur{Johnson}}

\affil[2]{\orgdiv{Department of Mathematics}, \orgname{Massachusetts Institute of Technology}, \orgaddress{\street{77~Massachusetts~Ave.}, \city{Cambridge}, \postcode{02139}, \state{MA}, \country{USA}}}

\affil*[1]{\orgdiv{Institute for Soldier Nanotechnologies}, \orgname{Massachusetts Institute of Technology}, \orgaddress{\street{77~Massachusetts~Ave.}, \city{Cambridge}, \postcode{02139}, \state{MA}, \country{USA}}}

\abstract{
A key challenge in topology optimization (TopOpt) is that manufacturable structures, being inherently binary, are non-differentiable, creating a fundamental tension with gradient-based optimization. The subpixel-smoothed projection (SSP) method~\citep{hammond_unifying_2025} addresses this issue by smoothing sharp interfaces at the subpixel level through a first-order expansion of the filtered field. However, SSP does not guarantee differentiability under topology changes, such as the merging of two interfaces, and therefore violates the convergence guarantees of many popular gradient-based optimization algorithms. We overcome this limitation by \emph{regularizing} SSP with the Hessian of the filtered field, resulting in a twice-differentiable projected density during such transitions, while still guaranteeing an almost-everywhere binary structure. We demonstrate the effectiveness of our second-order SSP (SSP2) methodology on both thermal and photonic problems, showing that SSP2 has faster convergence than SSP for \emph{connectivity-dominant} cases---where frequent topology changes occur—while exhibiting comparable performance otherwise. Beyond improving convergence guarantees for CCSA optimizers, SSP2 enables the use of a broader class of optimization algorithms with stronger theoretical guarantees, such as interior-point methods. Since SSP2 adds minimal complexity relative to SSP or traditional projection schemes, it can be used as a \emph{drop-in replacement} in existing TopOpt codes.}


\maketitle

\section{Introduction}\label{sec:introduction}

In density-based topology optimization (TopOpt), projection methods play a central role because they ensure that the optimized structure is binary and, therefore, manufacturable~\citep{wang_projection_2011}. In a typical three-field TopOpt formulation~\citep{sigmund_topology_2013}, projection is the final step, following the smoothing of the design density---the latter representing the optimization degrees of freedom (DOFs).
The main role of the projection operation is to drive the filtered field toward the extreme values of zero and one, representing the void and solid regions, respectively, in two-phase TopOpt. A practical challenge is to reconcile binary projection with differentiability, including a well-conditioned second-derivative, to ensure fast convergence with gradient-based optimization algorithms.

A common projection method employs the hyperbolic tangent function, where the steepness parameter $\beta \in [0, \infty)$ controls the projection strength: for $\beta = 0$, the projected field coincides with the filtered field, while for $\beta \to \infty$, the field becomes fully binarized and thus manufacturable. However, for large $\beta$, the Hessian of the projected field becomes ill-conditioned, which can slow down or even stall gradient-based optimization~\citep{nocedal2006numerical}. To mitigate this issue, a common strategy is to start with a weak projection and gradually increase $\beta$ as the optimization progresses~\citep{sigmund_topology_2013}. Although this continuation approach enables rapid early iterations, it may still result in slow convergence as $\beta$ grows. In practice, though, the final $\beta$ remains finite, and full binarization is achieved through postprocessing that hopefully does not perturb the structure too far from optimality. The choice of $\beta$-scheduling  involves a trade-off between computational efficiency and the quality of the final binarized structure. Recent studies have explored automatic $\beta$-scheduling~\citep{dunning_automatic_2025}, albeit at the cost of additional hyperparameters.  Although  many authors have learned tractable $\beta$  schedules for a wide variety of practical problems, especially for problems where the physics inherently favors the binary extremes, we have recently shown that the process can be dramatically simplified by using a new projection. The subpixel-smoothed projection (SSP)~\citep{hammond_unifying_2025} algorithm addresses this problem by locally smoothing the projected field using \emph{both} the filtered field and its spatial gradient, which together enable the local construction of an approximate \emph{signed distance} to the interface. This quantity is then used to smooth the projected field over a region roughly one pixel wide ($\Delta x$) around the interface, which remains twice differentiable even for $\beta = \infty$, while ensuring a strictly binary structure ``almost everywhere''---except within $\Delta x$ of interfaces---a property which we refer to in this paper as \emph{quasi-binary}. This feature enables simpler $\beta$ scheduling (or even optimizing with $\beta=\infty$ from the start) and faster convergence at high~$\beta$. One shortcoming of SSP is that it still yields nondifferentiable projected fields at $\beta=\infty$  as two interfaces merge~\citep{hammond_unifying_2025}, violating convergence guarantees of many optimization algorithms when such a common topology change occurs, motivating continued use of $\beta$ scheduling to discover the correct topology before (quasi-)binarizing. In fact, optimizers that are widely used in TopOpt, such as those based on the conservative convex separable approximation (CCSA)~\citep{svanberg_class_2002}, including the globally convergent MMA algorithm, and interior-point methods~\citep{wachter2006implementation} require in their convergence proofs that the cost function be twice-differentiable, though we find in practice (without proof) that they \emph{may} tolerate occasional discontinuities. 

To address this limitation, we expand the filtered field to second order and use its Hessian to \emph{regularize} the SSP signed distance. This methodology, which we term “second-order SSP’’ (SSP2), ensures that the projected density remains twice differentiable as two interfaces merge.  For clarity throughout the text, we refer to the original SSP as \emph{SSP1}. Importantly, when interfaces are well separated, SSP2 recovers the behavior of SSP1, preserving SSP1's central property of quasi-binary and differentiable structures for $\beta = \infty$.

We first apply SSP2 to a simple 1D analytic example based on a parametrized quadratic filter, which represents two separated interfaces with tunable distance. This example illustrates several regimes as a function of the interface separation and highlights the improved smoothness properties of SSP2 compared to SSP1. We then consider three TopOpt problems, running 100 optimizations with random initialization in each case. Notably, all optimizations begin directly from quasi-binary structures, corresponding to $\beta = \infty$. The first example is the design of a thermal metamaterial in which one phase has negligible thermal conductivity. Here, SSP2 exhibits overall faster convergence than SSP1, a behavior we attribute to the connectivity-dominant nature of the problem, which induces frequent topology changes. Remarkably, SSP1 still converges in most cases---though fewer than SSP2---despite the discontinuities introduced by topology transitions and the resulting lack of theoretical convergence guarantees. Using the converged structures as initial guesses, we subsequently impose a minimum feature size~\citep{zhou_minimum_2015} and find that the two methods achieve comparable performance. We attribute this behavior to the fact that the optimized constrained structures have topologies similar to their initial guess. The second example concerns the same thermal problem, but with the two phases having comparable thermal conductivity; because both phases support heat transport, optimal structures can include isolated islands, therefore, the problem is not connectivity-dominated. Consequently, for unconstrained optimization, both SSP1 and SSP2 converge for all samples with similar performance. The third example is the design of a wavelength demultiplexer. In this photonic TopOpt problem, the phases have similar dielectric permittivity, and, consistent with the second thermal case, SSP1 and SSP2 display comparable convergence behavior.

In practice, the Hessian is computed using bicubic interpolation and automatic differentiation (AD), thus introducing minimal complexity to the original SSP, making it a convenient \emph{drop-in replacement} in existing topology optimization workflows. SSP2 is easy to use, no slower to converge than SSP1 and sometimes much faster, and importantly yields much stronger theoretical convergence guarantees for popular optimization algorithms---hence, we believe it should be an attractive default choice for many TopOpt problems.
\\

The paper is organized as follows. Section~\ref{sec:ssp_overview} reviews the three-field formulation of TopOpt and the SSP1 method. Section~\ref{sec:ssp2} introduces SSP2 and illustrates its differentiability with two synthetic quadratic filtered fields. A geometric interpretation of SSP2 is presented in Appendix~\ref{sec:distance}. Section~\ref{sec:examples} reports the three realistic TopOpt examples, and conclusions are given in Section~\ref{sec:conclusion}.

\section{SSP1 review}\label{sec:ssp_overview}

In the three-field approach~\citep{sigmund_topology_2013},   TopOpt is performed in terms of the design field $\rho(\mathbf{x}) \in \mathcal R^n\rightarrow \mathcal R$, ($n$ being the dimensionality of the problem), which continuously varies in $[0,1]$. To avoid checkerboard patterns, this density undergoes filtering, often with the conic filter
\begin{equation}
\kappa_c(\mathbf{x}) =
\begin{cases}
   a \left(1-\frac{|\mathbf{x}|}{\tilde R}\right) , & |\mathbf{x}| \leq \tilde R, \\
  0,   & \text{otherwise},
\end{cases}
\end{equation}
where $\tilde R$ is the radius of the kernel, and $a$ defined so that $\int_{\Omega}\kappa_c \,d\mathbf{x} = 1$. The filtered field, $\tilde \rho = \kappa*\rho$, is then projected $\hat \rho (\tilde\rho)$ to ensure a final binary structure. A common projection function is  
\begin{equation}\label{eq:projection}
 \hat \rho_t =  \frac{\tanh{\beta \eta} + \tanh\left(\beta\left(\tilde{\rho}-\eta\right)\right)}{\tanh{\beta \eta}+ \tanh\left(\beta\left(1-\eta\right)\right)},
\end{equation}
where $\eta = 0.5$ defines the level set and $\beta \in [0^+,\infty)$ is the steepness parameter. As discussed in Sec.~\ref{sec:introduction}, $\beta$ typically increases gradually during optimization so that the optimizer navigates various topologies before the structure becomes near-binary. However, the key problem with this approach is that for high $\beta$  the cost function becomes stiff due to the ill-conditioned Hessian of the projected field, potentially stalling optimization~\citep{hammond2022multi}.  This is especially problematic in physical problems that inherently favor grayscale structures, in which for any finite $\beta$  it will still converge to a non-binary structure (via $\rho$ values near $\eta$) without additional penalties or constraints. The subpixel-smoothed projection (SSP1) avoids this problem~\citep{hammond_unifying_2025},  by analytically smoothing only the void-to-solid transition, which is designed to occur within a given \emph{smoothed} radius $\hat R \approx 0.5 \Delta x\ll \tilde R$. Concretely, let $\tilde \rho = \eta$ be the level-set defining a solid-void interface. We perform a first-order Taylor expansion of the filtered density,
\begin{equation}\label{eq:expansion}
\tilde \rho(\mathbf{x}') = \tilde \rho(\mathbf{x}) + \boldsymbol \nabla \tilde \rho (\mathbf{x}) \cdot \left(\mathbf{x}'-\mathbf{x}\right),
\end{equation}
from which we approximate the \emph{signed} distance from the interface
\begin{equation}\label{eq:distance}
 d_1(\mathbf{x}) = \frac{\eta - \tilde \rho(\mathbf{x})}{||\boldsymbol \nabla \tilde \rho(\mathbf{x})||}.
\end{equation}
 This distance is positive in the void region and negative in the solid region. Throughout the text, we use the notation $\nabla \tilde\rho (\mathbf{x}) = \left.\nabla \tilde \rho \right|_{\mathbf{x}}$, often omitting the argument $\mathbf{x}$. We define the ``fill-factor''
 \begin{equation}
FF_s(a,b)  =  \left[1 - F(s)\right]a + F(s)b,
\end{equation}
with 
\begin{equation}\label{eq:F2}
F(s) = \tfrac{1}{2}
 - \tfrac{15}{16}s
 + \tfrac{5}{8}s^{3} 
 - \tfrac{3}{16}s^{5},
\end{equation}
a polynomial designed to be twice differentiable when it is piecewise set to $1$ for $s < -1$ and to $1$ for $s > 1$. Then, the SSP projection is~\citep{hammond_unifying_2025}
\begin{equation}\label{eq:ssp}
\rone(\mathbf{x}) =
\begin{cases}
FF_{\hat d}(\hat \rho^-,\hat \rho^+), & \text{if } |\hat d| < 1, \\
\hat \rho, & \text{otherwise,}
\end{cases}
\end{equation}
where $\hat d = d/\hat R$ and $\hat \rho^{\pm}=\hat \rho(\tilde \rho^{\pm})$, with
\begin{equation}\label{eq:pm}
\tilde\rho^\pm_1(\mathbf{x}) =
\tilde{\rho}(\mathbf{x}) \pm  \hat{R}F(\mp \hat d_1) \, \|\boldsymbol \nabla \tilde{\rho}\| \,.
\end{equation}
The functions $\tilde \rho_{\pm}$ are the projection of the linear approximation (based only on the local properties $\tilde \rho (\mathbf{x})$ and $\boldsymbol \nabla \tilde \rho$) onto two points on either side of the level set, i.e., $-\hat R F(\hat d)$ and $\hat R F(-\hat d)$. Using the relationship $F(s) = 1- F(-s)$, it is straighforward to prove that the distance between these two points is $\hat R$. The key advantage of SSP becomes apparent for $\beta = \infty$. In this regime, Eqs.~\ref{eq:ssp}-\ref{eq:pm} become
\begin{equation}\label{eq:ssp_infty}
\hat\rho_1(\mathbf{x}) =
\begin{cases}
\begin{aligned}
F(\hat d_1)
\end{aligned}
& |\hat d_1|\leq 1, \\[0.8em]
H(-\hat d_1), & \text{otherwise},
\end{cases}
\end{equation}
with $H(s)$ being the Heaviside function. Because $F(s) = H(-s)$ and $F'(s) = 0$ for $|s|\to 1^-$, the transition of the projected field across the interface is twice-differentiable and binary except within $\hat R$ of the interface.

\section{Second-order SSP}\label{sec:ssp2}
Although SSP1 yields a twice-differentiable projected field across interfaces in the limit $\beta = \infty$, it does not guarantee smoothness when interfaces merge~\citep{hammond_unifying_2025}. In fact, in the middle point, here denoted with $\mathbf{\bar{x}}$, the gradient of the filtered field $\boldsymbol \nabla \tilde \rho(\mathbf{\bar x}) = 0$, leading to $|\hat d_1|=\infty$. The corresponding projected density at $\mathbf{\bar{x}}$ (Eq.~\ref{eq:ssp_infty}) is
\begin{equation}
\rone(\mathbf{\bar{x}})=H(\tilde \rho(\mathbf{\bar{x}})-\eta).
\end{equation}
 As the two interfaces approach each other, the value of $\tilde\rho(\bar{\mathbf{x}})$ increases, and when it crosses $\eta$ it triggers a discontinuity in $\hat \rho_1$. This behavior challenges the convergence of  optimizers such as CCSA, where the cost function is theoretically assumed to be twice-differentiable~\citep{svanberg_class_2002}. We tackle this problem by adding a space-dependent regularizer to the SSP1 distance (Eq.~\ref{eq:distance}), 
 \begin{equation}\label{eq:dSSP2}
d_2(\mathbf{x})
= \frac{\eta - \tilde\rho(\mathbf{x})}
       {\sqrt{\,||\boldsymbol\nabla \tilde\rho(\mathbf{x})||^2
       + \hat R^{2}\,\lVert\mathbf H(\mathbf{x})\rVert_{F}^{2}} },
\end{equation}
where $\mathbf H(\mathbf{x})$ is the Hessian of the filtered field and $|| \cdot||_F$ is the Frobenius norm. As this method is based on a second-order approach, we term it ``second-order smoothed-subpixel projection'' (SSP2), and refer to the corresponding distance (Eq.~\ref{eq:dSSP2}) as $d_2$. Within SSP2, at $\mathbf{x}=\bar {\mathbf{x}}$, we have
\begin{equation}\label{eq:dSSP20}
d_2(\mathbf{\bar{x}})
= \frac{\eta - \tilde\rho(\mathbf{\bar{x}})}
      { \hat R\,\lVert\mathbf H(\mathbf{\bar{x}})\rVert_{F}},
\end{equation}
which leads to a twice-differentiable projection. Similarly to SSP1, the requirement $\tilde R \gg \hat R$ is motivated by the need to ensure the validity of the quadratic approximation.
In regions away from $\mathbf{\bar x}$, SSP2 recovers the SSP1 limit, because the first-order term in the Taylor series dominates the second-order term except when the first-derivative is small. 
In fact, as derived in Sec.~\ref{sec:distance}, for quadratic fields and $n$-dimensional problems Eq.~\ref{eq:dSSP2} can be approximated by
\begin{equation}\label{eq:d4}
d_2(\mathbf{x}) \approx d_1(\mathbf{x}) \frac{1}{\sqrt{n\left(\frac{\hat R}{||\mathbf{x}-\mathbf{\bar x}||}\right)^2+1}},
\end{equation}
which clearly shows that the SSP1 distance is recovered when $||\mathbf{x}-\bar{\mathbf{x}}||\gg\sqrt{n}\hat R$. Additionally, in the limit $\hat R \to 0$, the two projections coincide; consequently, SSP2 preserves the property of yielding quasi-binary fields.


To generalize SSP2 to finite $\beta$, we also regularize Eq.~\ref{eq:pm} with the Hessian of the filtered field, 
\begin{equation}\label{eq:pm2}
\tilde\rho^\pm_2(\mathbf{x}) =
\tilde{\rho}(\mathbf{x}) \pm  \hat{R}F(\mp \hat d_2)\sqrt{\|\boldsymbol \nabla \tilde{\rho}\|^2+\hat R^2\lVert\mathbf{H}\rVert_F^2}  ,
\end{equation}
where $\hat d_2 = d_2/\hat R$. The SSP2 projection for any~$\beta$ becomes
\begin{equation}\label{eq:ssp2}
\hat{\rho}_2(\mathbf{x}) =
\begin{cases}
FF_{\hat d_2}(\hat \rho^-,\hat \rho^+ ), & \text{if } |\hat d_2| < 1, \\\hat \rho, & \text{otherwise,}
\end{cases}
\end{equation}
where $\hat \rho^{\pm}=\hat \rho(\tilde \rho_2^{\pm})$, with $\tilde \rho^{\pm}_2$ given by Eq.~\ref{eq:pm2}. Taken together, SSP2 only requires two modifications to SSP1, namely substituting Eq.~\ref{eq:d1} with Eq.~\ref{eq:d2} and Eq.~\ref{eq:pm} with Eq.~\ref{eq:pm2}, with the rest of the formalism being unaltered. In practice,  the terms $\boldsymbol \nabla \tilde \rho$ and $\mathbf{H}$ are calculated easily on regular grids by automatic differentiation and bicubic interpolation, using the JAX package Interpax~\citep{interpax2025}.  (In a finite-element method on an unstructured mesh, one could instead use second-order elements to represent $\tilde \rho$ and hence have access to its second derivatives.) In this paper, the gradient used in SSP1 is also computed via bicubic interpolation, unlike the original implementation based on linear interpolation~\citep{hammond_unifying_2025}. This ensures that performance differences in our comparison of the the two projections arise solely from regularization effects. Owing to its minimal added complexity, SSP2 can straightforwardly be used as a \emph{drop-in} replacement; note that nothing about the physics solver depends on the details of the projection algorithm.

\begin{figure*}[h!]
     \centering
     \includegraphics[width=1\textwidth]{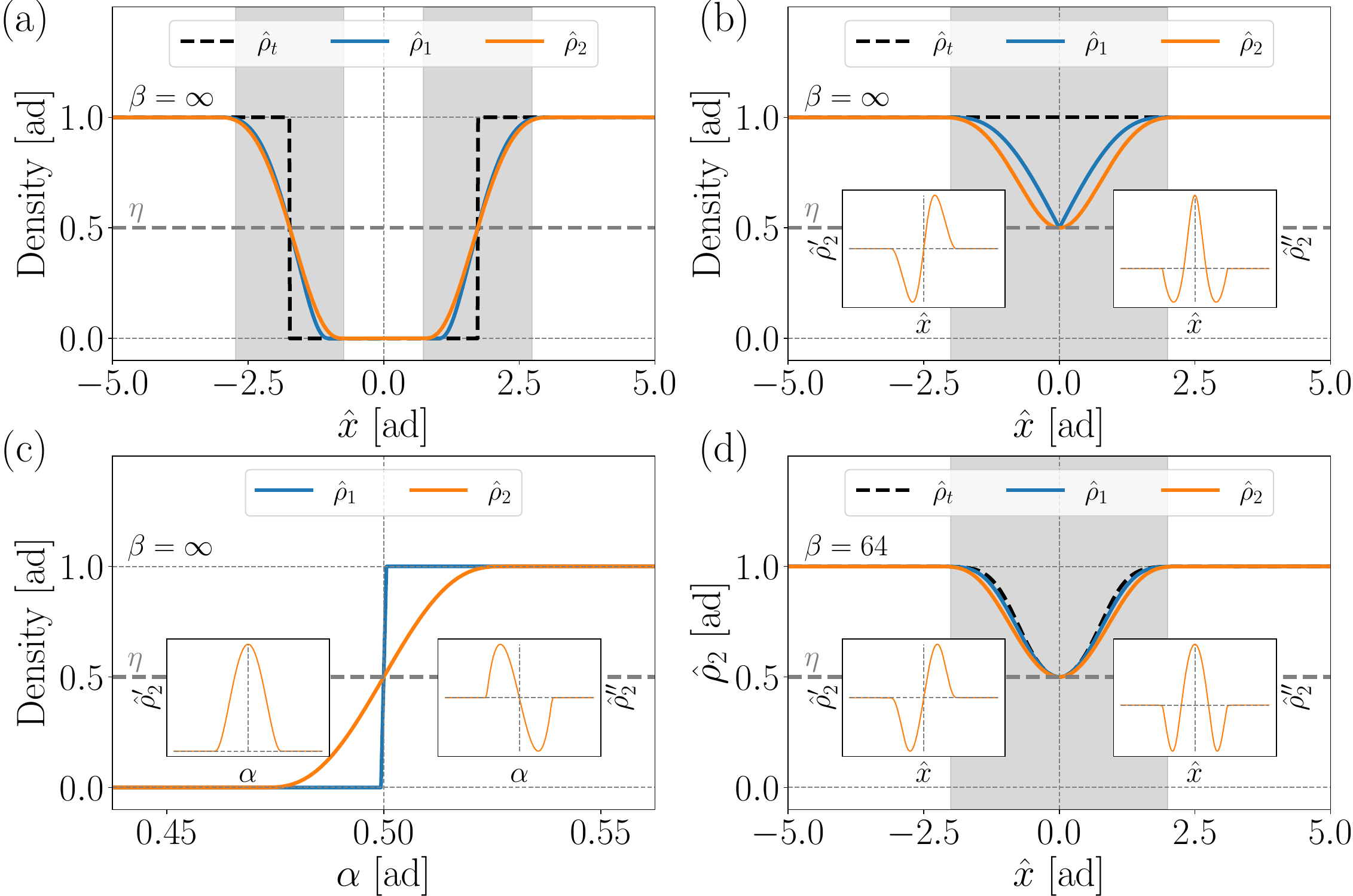}
     \caption{(a) The projection for $\beta = \infty$ and far interfaces. (b) The projections for $\beta = \infty$ and $\alpha = \eta$. The insets show the (left) first and (right) second derivative of $\rtwo$, (c) A comparison between $\rone$ and $\rtwo$ versus $\alpha$. At $\alpha=\eta=0.5$, $\rone$ presents a discontinuity. The insets show the (left) first and (right) second derivative of $\rtwo$. (d) The projections for $\beta=64$ and $\alpha = \eta$. The insets show the (left) first and (second) derivative of the SSP2 projection. In (a), (b) and (c), the $x$-axis is the normalized position $\hat x  = x/\hat R$.}
     \label{fig1}
\end{figure*}

\subsection{Parabolic field}
To  illustrate the SSP2 methodology, we first consider a synthetic case with $\beta = \infty$ and a parabolic filtered field
\begin{equation}\label{eq:expansion_parabolic}
\tilde \rho(x) = \alpha + \frac{1}{2}\left(\frac{x}{\tilde R}\right)^2,
\end{equation}
where $\tilde R = 6\hat R$. The projection of this field corresponds to two parallel interfaces whose separation is parametrized by $\alpha$. For values of $\alpha < \eta$ such that the distance between the interfaces is much larger than $\hat R$, SSP2 recovers SSP1, i.e., $\rtwo \approx \rone$. (Fig.~\ref{fig1}(a)). The precise value of $\alpha$ at which the SSP1 transition occurs is derived in the Supplemental Information. For $\alpha = \eta$, $\rone(0) = \eta$ but it is nondifferentiable at the origin (Fig.~\ref{fig1}(b)), whereas $\rtwo$ is twice differentiable (inset of Fig.~\ref{fig1}(a)). This regime corresponds to the case where the tanh projection ($\hat \rho_t$) becomes completely solid. In addition to the non-differentiability in $x$ at $(\alpha = \eta,x=0)$, SSP1 also exhibits a discontinuity in $\alpha$ at the same point. In fact, $\rone(0,\alpha) = H(\alpha-\eta)$ (Fig.~\ref{fig1}(c)). In contrast, $\rtwo(0)$ is twice-differentiable in $\alpha$ (insets of Fig.~\ref{fig1}(c)). Taken together, these results illustrate how SSP2 provides a smooth response to changes in topology, while still remaining quasi-binary. The positivity of the regularization enforces \(\hat d_2 \le \hat d_1\) and therefore reduces the frequency of fallbacks to the \(\tanh\) projection (see Eq.~\ref{eq:ssp2}); as shown in the Supplemental Information, for a parabolic field this mechanism manifests as the need for a larger value of \(\alpha\) to merge two approaching features compared to SSP1. Nevertheless, in our experiments, we observe that the additional grayness introduced by SSP2 is negligible ($< 1\%$), and in any cases the gray regions near interfaces still vanish with increasing resolution ($\Delta x \to 0$).

For finite values of $\beta$, the two projections are practically identical. Figure~\ref{fig1}(d) compares the projection methods for $\beta = 64$ and $\alpha=\eta$, where we have $\rone\approx\rtwo\approx\hat \rho_t$. Lastly, $\rtwo$ is twice-differentiable throughout the domain, illustrating how Eq.~\ref{eq:ssp2} ensures smoothness as SSP2 transitions to $\hat \rho_t$ in $|\hat d_2| = 1$ (insets of Fig.~\ref{fig1}(d)). 

\subsection{Cassini ovals}
To further illustrate the effect of SSP2, we also considered a two-dimensional case in which the level set $\tilde \rho = \eta$ is represented by the Cassini oval~\citep{lawrence2013catalog}---a simple closed polynomial (quartic) curve that exhibits a topology change as one of its parameters is varied.  Unlike the previous example, the topology now changes by two regions meeting at a single \emph{point}, rather than two parallel flat interfaces colliding.  The filtered field is given by
\begin{equation}\label{eq:cassini}
\begin{aligned}
\tilde \rho(x,y,b)
&= \bigl[(x-a)^2 + y^2\bigr]\bigl[(x+a)^2 + y^2\bigr] -\\
&\quad - b^{4} + \eta.
\end{aligned}
\end{equation}
Here, the parameter $a$ (which we fix at $a=1$) determines the location of the foci of two \emph{blobs}, located at $(-a,0)$ and $(a,0)$, whereas the parameter $b^2$ is the \emph{focal product}, which determines the blobs size. The topology is parametrized by $e \coloneqq \left | b/a \right|$: for $e < 1$, the two blobs are separated; at $e = 1$, the two domains are in contact only at the point $(0,0)$; and for $e > 1$, the blobs merge into a single connected domain. To assess the smoothness properties of the projections, we evaluate the SSP1 and SSP2 projections at $(0,0)$ as a function of $d$. Similarly to the parabolic case, we find that SSP1 exhibits a jump discontinuity at $e = 1$, whereas SSP2 is twice differentiable. In the Supplemental Information, we include an expanded section on the case of the Cassini ovals.

\section{Examples}\label{sec:examples}

We apply SSP2 to two thermal problems and one Maxwell example. In all cases, the response of the physical system is determined by $M$ simulations, solved for various excitations. The projected density is used directly by the cost function $f$, the PDE solver, and the inequality constraints $g_c$ with $c=0,...,C-1$. In practice, the densities are discretized into a grid of $N$ pixels; using their vectorial notation, the optimization algorithm is
\begin{align}\label{eq:algo}
\min_{\boldsymbol \rho} \quad & f(\boldsymbol {\hat \rho}, \xi_0, \dots, \xi_{M-1}) \\
\text{s.t.} \quad 
& \mathcal{R}(\boldsymbol {\hat \rho}, \xi_m) = 0,     && m = 0, \dots, M - 1, \nonumber\\
\text{s.t.} \quad 
& g_c(\boldsymbol {\hat \rho}) \le 0,                  && c = 0, \dots, C - 1, \nonumber\\
\text{s.t.} \quad 
& 0 \le \rho_n \le 1,                                  && n = 0, \dots, N - 1, \nonumber
\end{align}
where we describe the PDE in its residual form, parametrized by the physical variable $\xi$. Equation~\ref{eq:algo} is solved using the quadratic-model variant of the
CCSA optimizer~\citep{svanberg_class_2002}, implemented in the open-source software \texttt{NLopt}~\citep{johnson2014nlopt}. 

\begin{figure*}[htbp]
     \centering
     \includegraphics[width=1\textwidth]{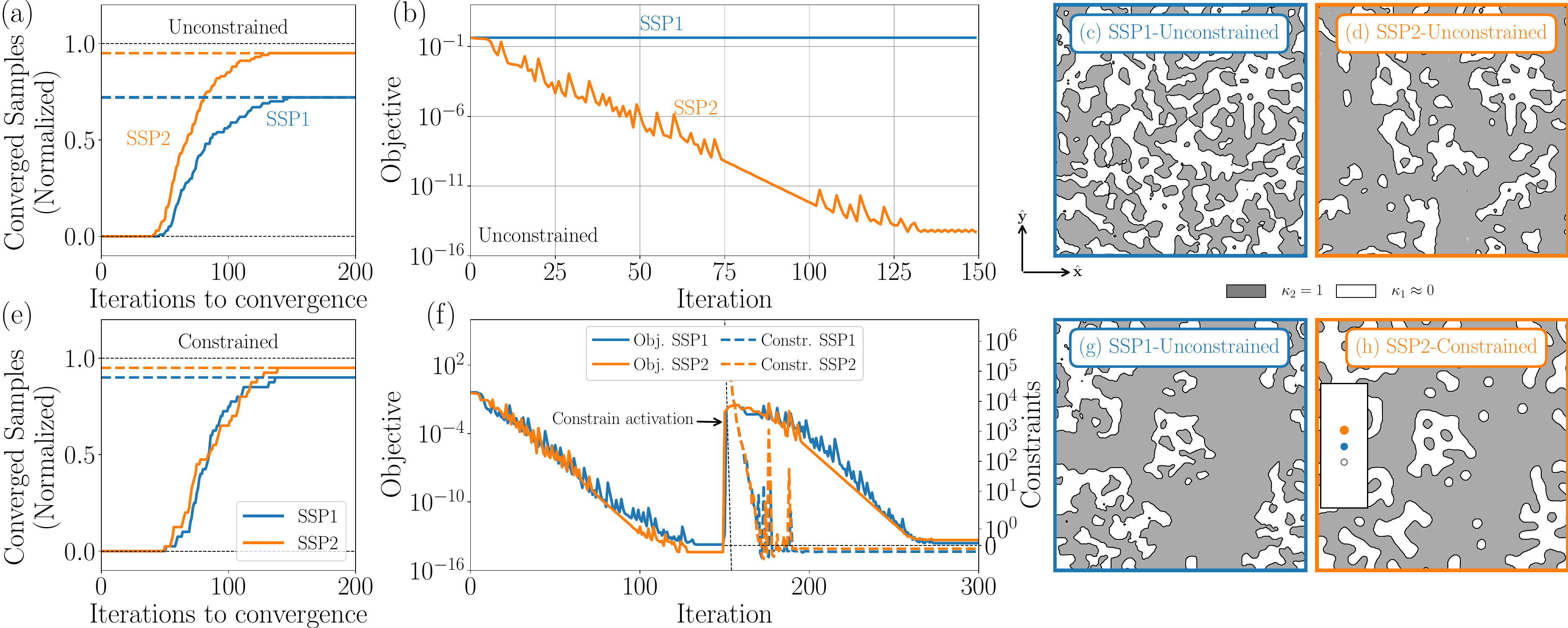}
     \caption{(a) The cumulative convergence ratio for unconstrained optimization, i.e., the number of converged cases over the total number of samples (100). A case is defined ``converged'' if the loss function falls below $10^{-7}$ within 150 steps. (b) Convergence for an example where SSP2 converges and SSP1 does not, and the corresponding optimized structures for (c)  SSP1 and (d) SSP2. (e) The cumulative convergence ratio when lengthscale are activated, with the total number of samples of 40. (f) Convergence for an exemplary case spanning both the unconstrained and constrained phase. The constraint function is also shown. (e) The first guess used for constrained optimization for both projections, and (h) the optimized structure obtained by SSP2 (SSP1 gives a nearly identical structure). The inset shows the (top) imposed lengthscales (for both the void and solid phase), and the obtained one for solid (center) and void (phase) phase.}
     \label{fig2}
\end{figure*}

\subsection{Thermal metamaterial optimization}\label{sec:thermal}
The first problem is the design of a 2D thermal metamaterial. Specifically, given a set of two materials with thermal conductivity  $\kappa_0\kappa_1\mathbf{\mathcal I}$ and $\kappa_0\kappa_2\mathbf{\mathcal I}$ ($\mathbf{\mathcal I}$ being the 2D identity matrix), the task is to identify the material distribution that leads to a target effective thermal conductivity, $\kappa_D=\kappa_0\bar{\kappa}$. We set $\kappa_0 = 1$Wm$^{-1}$K$^{-1}$. (due to linearity, the optimal structures are independent of $\kappa_0$) The first phase has negligible thermal conductivity $\kappa_1\approx0$, (in practice, we use a small number for regularization), while the second one has $\kappa_2=1$. We refer to this case as the ``porous'' regime. The normalized target effective conductivity is
\begin{equation}\label{eq:target}
\bar \kappa = \begin{pmatrix}
0.2 & 0 \\
0 & \bar 0.4
\end{pmatrix}.
\end{equation}
To reconstruct the effective thermal for a given structure, we solve the heat conduction equations with periodic boundary conditions and for three different temperature perturbations~\citep{romano_inverse_2022}. Thus, following the notation from Eq.~\ref{eq:algo}, we have
\begin{equation}
\mathcal{R} = \nabla \cdot \left[\kappa_0\kappa(\hat \rho) \nabla T_m\right], 
\quad m = 0, \dots, 2,
\end{equation}
where $\kappa(\hat \rho)$ is given by
\begin{equation}\label{eq:kappa}
\kappa(\mathbf{x}) = \kappa_1+ \hat{\rho}\left(\mathbf{x})  (\kappa_2 - \kappa_1\right).
\end{equation}
The expression of the cost function $g(\hat \rho,T_0,T_1,T_2)$ is reported in~\citep{arrieta2025hyperparameter}. 
\begin{table}[h!]
\centering
\begin{tabular}{lcc}
\hline
\textbf{Method} & \textbf{Unconstrained}  & \textbf{Constrained} \\
\hline
Both SSP1 and SSP2        & 71 & 34 \\
Neither method        & 4 & 0 \\
SSP2 only   & 24 & 4 \\
SSP1 only   & 1  & 2 \\
\hline
\end{tabular}
\caption{Count of successful convergence outcomes in 150 iterations for SSP1 and SSP2, with unconstrained (out of 100 samples) and constrained optimization (out of 40 samples).}
\label{table}
\end{table}
We discretized the heat conduction equation using the finite-volume method, where a square domain of side $L=1$ m is divided into $N=161\times 161$ volumes. The conic radious filter is $\tilde R = $ 5 pixels. We perform optimization for 100 random initial guesses, starting with $\beta = \infty$. To obtain a global measure of convergence, we define the normalized cumulative convergence as the fraction of samples that converge within a given number of iterations, normalized by the number of samples. A problem is considered converged if the loss function drops below $10^{-7}$ within 150 iterations. The results are summarized in Table~\ref{table}. As shown in Fig.~\ref{fig2}(a), both SSP1 and SSP2 converge in most cases, with SSP2 consistently achieving faster convergence. Specifically, 71 samples converged in 150 iterations for both cases, with 24 instances where only SSP2 succeeded. 
\begin{figure}[htbp]
     \centering
     \includegraphics[width=1\columnwidth]{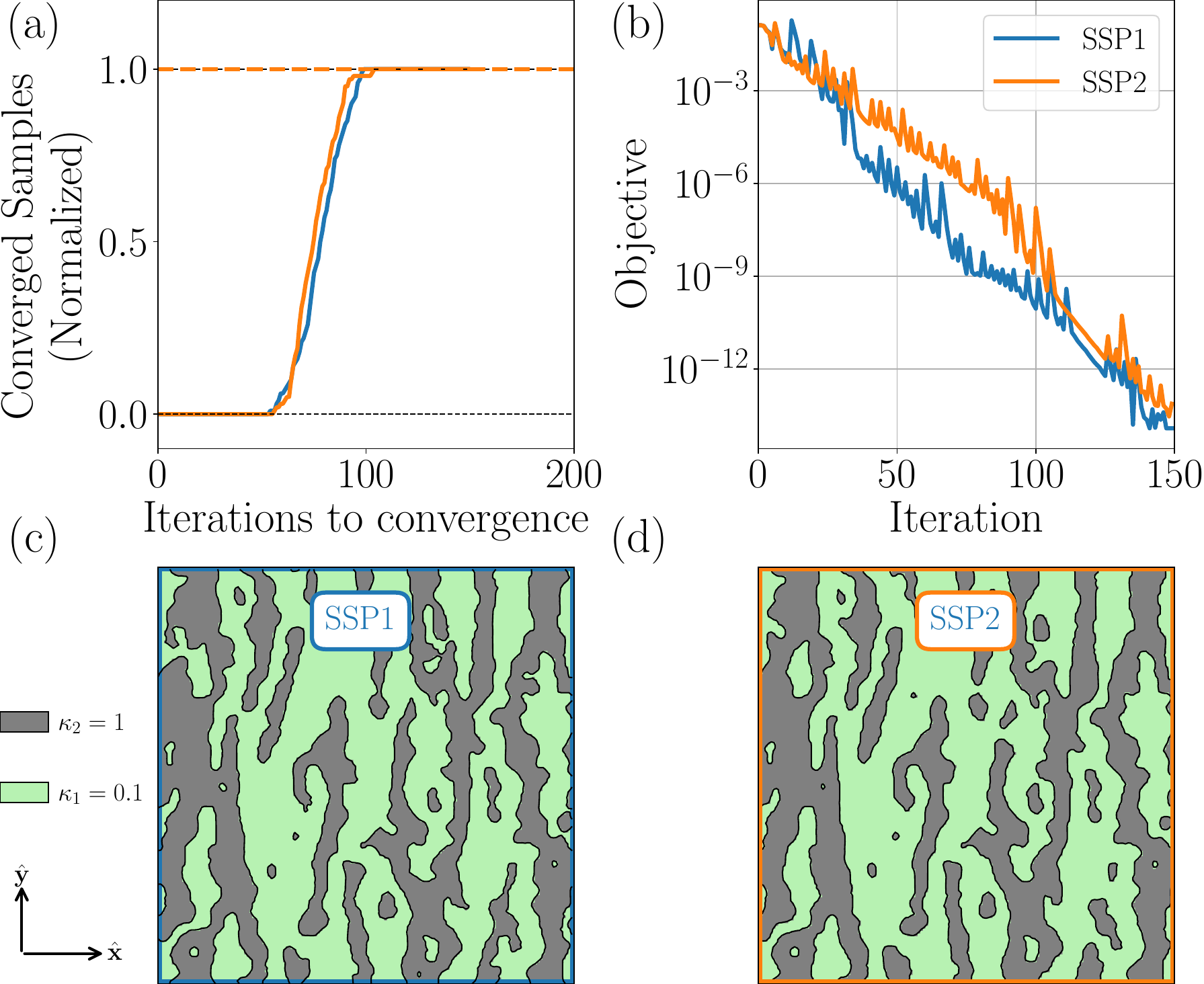}
     \caption{(a) Normalized cumulative convergence for a set of 100 optimizations of thermal transport in composite materials. Both SSP1 and SSP2 converge in all cases with comparable performance. (b) Example of convergence for both methods starting from the same random initial guess, and the corresponding optimized structures for the (c) SSP1 and (d) SSP2 cases.}
     \label{fig2b}
\end{figure}
Figure~\ref{fig2}(b) illustrates one such case: SSP1 stalls while SSP2 reaches a loss of about $10^{-15}$. The final structure obtained with SSP1 for this case (Fig.~\ref{fig2}(c)) consists of mostly disconnected regions, leading to zero heat flux and, consequently, vanishing effective thermal conductivity. To achieve the target value (Eq.~\ref{eq:target}), the optimizer must reconnect these regions, requiring a topology change. However, SSP1 struggles to restore connectivity, and the structure remains fragmented throughout optimization. In contrast, SSP2 smoothly transitions between topologies, eventually producing a satisfactory design (Fig.~\ref{fig2}(d)). In four cases, neither method converges, and, interestingly, in one instance SSP1 outperformed SSP2. In this case, SSP2 got stuck in a poor local minimum. Conversely, any  given algorithm (such as SSP1 in this case) can sometimes get “lucky” with its initial large steps and find a better minimum. This aspect motivated our statistical analysis, rather than relying on a single example, to assess the overall performance of SSP2.

We note that initializing the optimization directly at \(\beta=\infty\) leads to more frequent topology changes than starting from a finite \(\beta\), an effect that is especially pronounced when using random initialization, as in our approach. This scenario therefore represents the most favorable regime for SSP2. In practice, however, the standard $\beta$-progression strategy typically yields a structure that is already close to its final topology by the time $\beta=\infty$ is applied, which may reduce the performance gap between SSP1 and SSP2. On the other hand, these results suggest that a more aggressive $\beta$-schedule may be viable, since SSP2 at $\beta=\infty$ yields smoother transitions during topology changes. Investigating this balance in practical settings is a promising direction for future work.
\\

We now compare the two methods under minimum feature-size constraints for both phases. To this end, we adopt the geometric constraints~\citep{zhou_minimum_2015}, using the hyperparameters $c_0 = 64 \tilde R^2$ and $\epsilon = 10^{-8}$, which were derived from first principles~\citep{arrieta2025hyperparameter}. As these constraints act on the void and solid phases independently, we have two inequality constraints ($C = 2$ in Eq.~\ref{eq:algo}). From the set of cases that successfully converged with SSP1, we randomly select 40 and use them as initial guesses for both methods. This strategy ensures that SSP2 does not benefit from a more favorable initialization, although we found that for a given $\rho$, the corresponding $\rtwo$ typically remains very close to $\rone$, with relative differences in the loss function below $1\%$. The results of the constrained optimization are summarized in Table~\ref{table}. In contrast to the unconstrained case, the two methods exhibit comparable performance: in 24 instances, both SSP1 and SSP2 converge within 150 iterations; in four cases only SSP2 achieves convergence, whereas in two cases convergence is reached exclusively by SSP1. There are no instances in which both methods fail. Figure~\ref{fig2}(f) shows the convergence for a case where both methods converge for the initial guess (a converged SSP1 sample) illustrated in Fig.~\ref{fig2}(g). The two methods converge roughly at the same rate, with the constraints satisfied within the first 50 iterations in both cases. The corresponding optimized structures are practically the same. As shown in Fig.~\ref{fig2} the final design did not undergo significant topology changes compared to the initial guess, explaining the similar performance of the two methods in this case. For validation, we evaluate the lengthscale of the optimized structure using \texttt{imageruler} software~\citep{chen_validation_2024}, whose estimate is within $\pm$ 1 pixel~\citep{arrieta2025hyperparameter}. We obtain 4 pixels for both the solid and the void phases. Since the imposed lengthscale is 5 pixels, the optimized structure is successfully validated. 
Motivated by these results, we next consider the case in which the two phases have comparable relative thermal conductivity, specifically $\kappa_1 = 0.1$ and $\kappa_2 = 1$, and do not impose lengthscale constraints. In this case, which we refer to as the ``composite'' regime, both phases support non-negligible heat flow, and the optimization is therefore expected to be not connectivity-dominated, narrowing the performance gap between SSP2 and SSP1. To test this hypothesis, we perform 100 optimizations from different random initial conditions, analogously to the previous case. As anticipated, the two methods exhibit overall similar convergence behavior (Fig.~\ref{fig2b}(a)). Notably, unlike in the previous examples, all samples converge. To gain qualitative insight, we show the optimized structures for a representative realization obtained with SSP1 and SSP2 in Fig.~\ref{fig2b}(c) and Fig.~\ref{fig2b}(d), respectively. Both designs feature several isolated islands, consistent with topology changes such as interface merging being less prevalent than in the porous regime. Furthermore, we note that these islands are predominantly vertical, thereby favoring heat transport along the $y$-axis over the $x$-axis, consistent with the anisotropy of the target effective thermal conductivity (Eq.~\ref{eq:target}).

\subsection{Design of Wavelength Demultiplexer}\label{sec:photonics}

\begin{figure}[htbp]
     \centering
     \includegraphics[width=1\columnwidth]{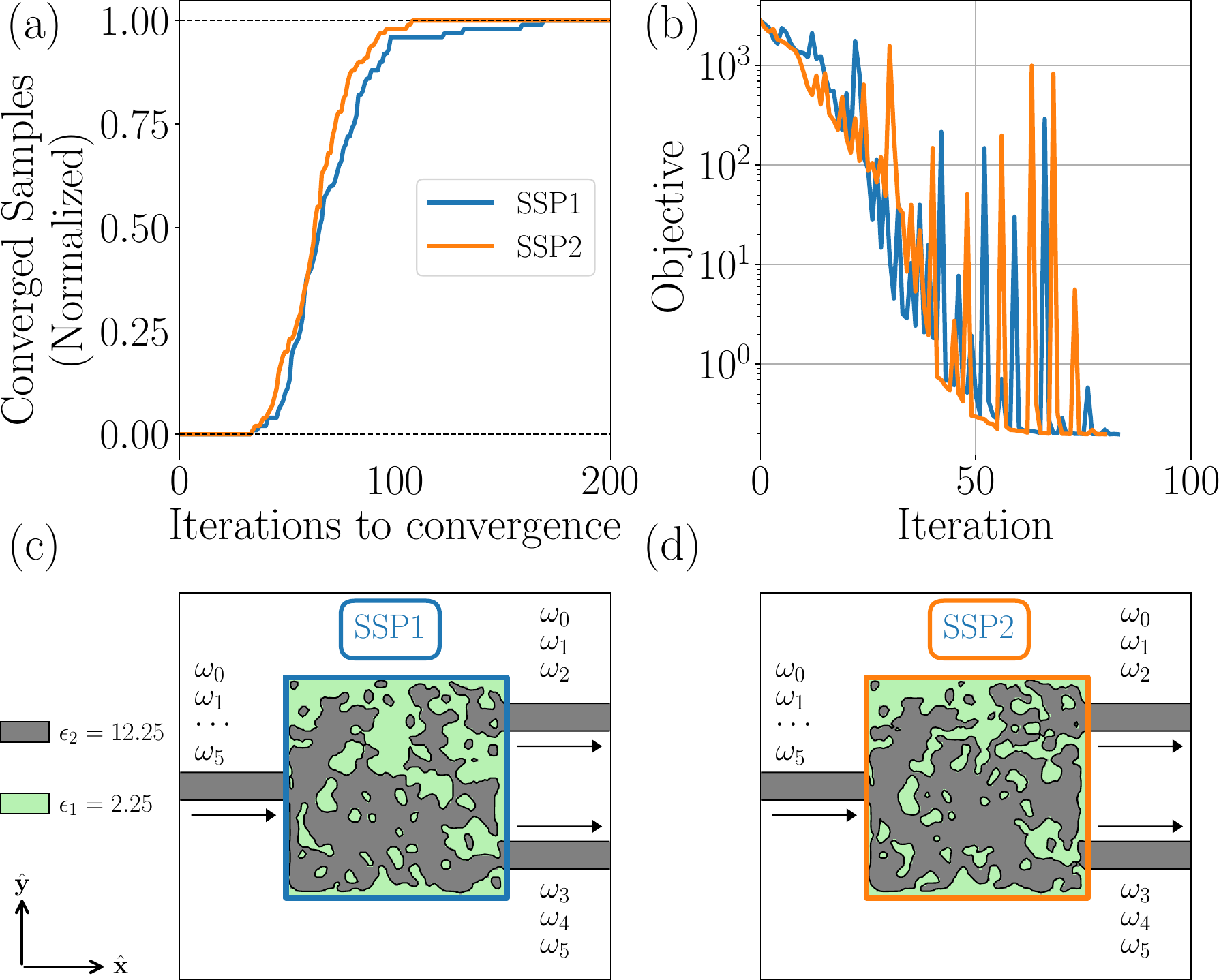}
     \caption{(a) The number of samples achieving a loss function $<1$ within a given number of iterations ($x$-axis) (b) An example of convergence trajectory for both approaches, using the same initial guess and their corresponding optimized structure for (c) SSP1 and (d) SSP2.}
     \label{fig4}
\end{figure}
In this section, we consider the design of a 2D photonic wavelength demultiplexer with no length-scale constraints. The device, detailed in~\citep{schubert2022inverse}, has one input port and two output ports, each connected to a semi-infinite, fixed waveguide of width $400\,\mathrm{nm}$. The design region is a $3.2 \times 3.2~\mu\mathrm{m}^2$ rectangle discretized into $80 \times 80$ pixels of size $\Delta x = 40\,\mathrm{nm}$ (Fig.~\ref{fig4}c). The two material phases are silicon, with relative dielectric permittivity $\epsilon_2 = 12.25\,\mathcal I$, and silicon dioxide, with $\epsilon_1 = 2.25\,\mathcal I$. The goal is to route six incoming excitations into the two output ports: the top port is designed to support wavelengths $\lambda_0 = 1265\,\mathrm{nm}$, $\lambda_1 = 1270\,\mathrm{nm}$, and $\lambda_2 = 1275\,\mathrm{nm}$, while the bottom port targets $\lambda_3 = 1285\,\mathrm{nm}$, $\lambda_4 = 1290\,\mathrm{nm}$, and $\lambda_5 = 1295\,\mathrm{nm}$. The design problem follows the topology-optimization formulation of Eq.~(\ref{eq:algo}), with $N = 6400$ design variables, $M = 6$ excitations, $C = 0$ constraints, and residuals defined by the frequency-domain Maxwell equations
\begin{eqnarray}\label{eq:maxwell}
\mathcal{R} = \nabla\times\tfrac{1}{\mu_0}\nabla\times\elf_m-\omega_m^2\epsilon_0\epsilon_r(\hat\rho)\elf_m+i\omega_m \bol J_m \nonumber
\end{eqnarray}
with $m=0,\dots,M-1$. In Eq.~\ref{eq:maxwell}, $\elf_m$ is the electric-field response for a given angular frequency $\omega_m$ and current density $\mathbf{J}_m$, $\mu_0$ is the vacuum  permeability, as $\epsilon_0$ is the vacuum permittivity. The local relative dielectric permittivity, $\epsilon_r$, depends on the position via the projected density. Similarly to the thermal case, its value is given by a linear interpolation between the values of the two phases, and is given by
\begin{equation}
\epsilon_r(\mathbf{x}) = \epsilon_1+ \hat{\rho}\left(\mathbf{x})  (\epsilon_2 - \epsilon_1\right).
\end{equation}
The definition of the cost function $g(\hat\rho,\mathbf{E}_0,\dots,\mathbf{E}_5)$ is provided in~\citep{schubert2022inverse}. The conic radious filter is $\tilde R = $ 3 pixels.
\\
 The Maxwell equations are solved using \texttt{invrs\_gym}~\citep{schubert_inverse_2022}, a wrapper to \texttt{Ceviche}, open-source 2D finite differences frequency domain solver (FDFD)~\citep{hughes2019forward}. Similarly to the thermal case, we perform 100 optimizations with different random initialization directly with $\beta=\infty$. For both SSP1 and SSP2, all samples converge to a loss value below~1, exhibiting comparable performance, as shown by the normalized cumulative convergence with threshold~1 (Fig.~\ref{fig4}(a)). Figure~\ref{fig4}(b) shows the convergence trajectory for a representative realization, where in both approaches the loss function reaches $\sim0.1$, consistent with the values reported in~\citep{arrieta2025hyperparameter} for a similar geometry. The optimized structures (Fig.~\ref{fig4}(c) and Fig.~\ref{fig4}(d) for SSP1 and SSP2, respectively) contain several isolated islands; hence, analogously to the composite case, this problem is not connectivity-dominated.

\section{Conclusions}\label{sec:conclusion}

We present the second-order, subpixel-smoothed projection (SSP2), which is designed to provide a twice-differentiable projected field as two interfaces merge, thus meeting the convergence guarantees of widely used optimizers such as those based on CCSA~\citep{svanberg_class_2002} and interior-point methods~\citep{wachter2006implementation}. The proposed methodology extends the original SSP method (referred to in this work as ``SSP1''), which ensures smoothness only for single-interface scenarios. The key aspect of SSP2 is the expansion of the filtered field to second-order, with the Hessian acting as a \emph{regularizer} to SSP1, the latter being based only on a first-order approximation. We apply SSP2 to three examples, including thermal and photonics problems, with optimization performed directly in terms of quasi-binary structures (obtained in the SSP framework with $\beta=\infty$). We find that SSP2 has better convergence properties than SSP1 in cases where merging interfaces are \emph{required} to ensure satisfactory structures. This is the case, as examined in this work, where a thermal metamaterial has one phase with negligible thermal conductivity: the other phase is responsible for ensuring connectivity in order to support nonzero flux. A prominent example in this category is structural optimization, which represents a possible future direction for the SSP2 projection. 

Interestingly, SSP1 converges in roughly two thirds of the cases, even though convergence guarantees are not met for merging interfaces. In practice, it seems to be ``stepping over'' the discontinuities caused by topology changes. When the thermal conductivities of the two phases are comparable, the structures do not need to be connected, and the two projection methods exhibit similar performances. Similar conclusions are drawn for the design of a mode demultiplixer, where the two phases have comparable dielectric permittivity. Another case where there is no significant performance gap is when the topology does not change significantly during optimization, as for the case with lengthscale constraints~\citep{arrieta2025hyperparameter}.  Nevertheless, the lack of theoretical convergence guarantees is troubling, and other optimization algorithms may be even less robust to discontinuities.

In the presence of topology changes at $\beta = \infty$, SSP2 enables much stronger convergence guarantees than SSP1 for common optimization algorithms that assume twice differentiability.  In practice, it is sometimes faster to converge than SSP1 and is at worst no slower.  Implementation-wise, it is a drop-in replacement for SSP2 that only requires the modification of a few lines of code, along with an interpolation scheme for $\tilde \rho$ (e.g.~bicubic) that provides spatial second derivatives.  Therefore, we believe that it should almost always be attractive to use SSP2 as a default choice of projection algorithm in density-based TopOpt.

\appendix

\section{Geometric Interpretation of SSP2}\label{sec:distance}
In this section, we provide a geometric interpretation of the SSP2 distance, $\hat d_2$ (Eq.~\ref{eq:dSSP2}). Let us  assume that we have two interfaces with middle point $\mathbf{\bar{x}}$, that is, the point where $\boldsymbol \nabla \tilde \rho = 0$. We build an ansatz for $\hat d_2$ with the requirement that $\hat{d}_2(\mathbf{x})\approx \hat{d}_1(\mathbf{x})$ when we are far from $\bar{\mathbf{x}}$, that is,  $\mathbf{x} : s(\mathbf{x}) = |\mathbf{x}-\bar{\mathbf{x}}| >> \bar s$, for $\bar s$ being a thresholding parameter. A suitable choice for such an ansatz is
\begin{equation}\label{eq:d1}
d_2(\mathbf{x}) =   d_1(\mathbf{x})\frac{1}{\sqrt{\left(\frac{\bar s}{s(\mathbf{x})}\right)^2+1}}.
\end{equation}
The goal is to determine $\bar s$ and an expression for $s(\mathbf{x})$ that depends only on local properties of $\tilde \rho$. To this end, we perform a second-order expansion of the filtered field around a generic point $\mathbf{x}$,
\begin{equation}\label{eq:quadraticnewton}
\begin{aligned}
\tilde \rho(\mathbf{x}') &\approx \tilde \rho(\mathbf{x}) 
+ \boldsymbol \nabla \tilde\rho(\mathbf{x}) \cdot \left(\mathbf{x}'-\mathbf{x}\right)+ \\
&\quad + \frac{1}{2}\left(\mathbf{x}'-\mathbf{x}\right)^{\!T}
\mathbf{H}(\mathbf{x})\left(\mathbf{x}'-\mathbf{x}\right).
\end{aligned}
\end{equation}
The minimum of this expansion, $\bar{\mathbf{x}}$, is obtained by solving $\boldsymbol \nabla_{\mathbf{x}'}\tilde\rho = 0$, yielding
\begin{equation}\label{eq:conditions}
\boldsymbol \nabla \tilde\rho(\mathbf{x}) + \mathbf{H}(\mathbf{x}) \left(\mathbf{x}'-\mathbf{x}\right) = 0,
\end{equation}
whose solution is
\begin{equation}\label{eq:solution}
 \bar{\mathbf{x}} = \mathbf{x} -\mathbf{H}^+\boldsymbol \nabla \tilde \rho,
\end{equation}
where ``$^+$'' stands for pseudoinverse. Generalized inversion is needed for cases where $\mathbf{H}$ is singular, such as the case of a 2D straight interface.
The corresponding distance from the middle point is 
\begin{equation}\label{eq:sep}
 s(\mathbf{x}) = ||\mathbf{x} - \bar{\mathbf{x}}|| = ||\mathbf{H}^+\boldsymbol \nabla \tilde \rho||.
\end{equation}
Including Eq.~\ref{eq:sep} in Eq.~\ref{eq:d1} gives
\begin{equation}\label{eq:d2}
d_2 =  d_1\frac{1}{\sqrt{\left(\frac{\bar s}{ |\mathbf{H}^+\boldsymbol \nabla \tilde \rho|}\right)^2+1}}
\end{equation}
While Eq.~\ref{eq:d2} is based on the \emph{exact} $s(\mathbf{x})$ for quadratic fields, it is not differentiable at $\bar{\mathbf{x}}$, i.e. when $\boldsymbol \nabla \tilde \rho = 0$. In fact, when recasted into
\begin{equation}\label{eq:d2plus}
 d_2= \frac{\eta - \tilde \rho}{\sqrt{||\boldsymbol \nabla \tilde \rho||^2 + \left(\frac{\bar s}{||\mathbf{H}^+\mathbf{u}||}\right)^2}},
\end{equation}
with $\mathbf{u} = \boldsymbol\nabla \tilde \rho/|\boldsymbol\nabla \tilde \rho|$, it becomes apparent that $\lim_{\mathbf{x}\to\bar{\mathbf{x}}}\hat d_2(\mathbf{x})$  depends on $\mathbf{u}$. To sidestep this issue, we build the \emph{isotropic} Hessian approximation
\begin{equation}\label{eq:fro}
\bar{\mathbf{H}} \approx \frac{1}{\sqrt{n}}\lVert\mathbf{H}\rVert_F \mathcal{\mathbf{I}},
\end{equation}
where $n$ is the dimensionality of the problem. Equation~\ref{eq:fro} has the property of preserving the Frobenium norm of the true Hessian. Including Eq.~\ref{eq:fro} in Eq.~\ref{eq:quadraticnewton} gives 
\begin{equation}
s(\mathbf{x})\approx \sqrt{n}||\boldsymbol \nabla \tilde \rho||\lVert\mathbf{H}\rVert_F^{-1},
\end{equation}
which leads to
\begin{equation}\label{eq:d3}
d_2=\frac{\eta -\tilde \rho}{\sqrt{||\boldsymbol \nabla \tilde \rho||^2 + \frac{1}{n}{\bar s}^2\lVert\mathbf{H}\rVert_F^2 }}.
\end{equation}
For 2D systems, Eq.~\ref{eq:d3} recovers Eq.~\ref{eq:distance} for $\bar s = \hat R\sqrt{2}$; the SSP1 transition is best illustrated by rewriting Eq.~\ref{eq:distance} as
\begin{equation}\label{eq:d4}
d_2(\mathbf{x}) \approx d_1(\mathbf{x}) \frac{1}{\sqrt{2\left(\frac{\hat R}{||\mathbf{x}-\mathbf{\bar x}||}\right)^2+1}},
\end{equation}
which also shows that the two projections coincide in the limit $\hat R \to 0$; hence, analogously to SSP1, SSP2 yields quasi-binary fields.

\bmhead{Supplementary Information}
This work includes supplementary material.

\subsection*{Declarations}
\bmhead{Conflict of interest} The authors declare that they have no conflict of interest.

\bmhead{Funding}
This work was supported in part by the U.S. Army Research Office through the Institute for Soldier Nanotechnologies (Award No. W911NF-23-2-0121) and by the Simons Foundation through the Simons Collaboration on Extreme Wave Phenomena Based on Symmetries.

\bmhead{Ethics approval}
The authors declare that there are no ethical issues involved in this research.

\bmhead{Consent to participate}
The authors declare that they all consented to participate in this research.

\bmhead{Data availability}
All data are included in the article and its supplementary materials.

\bmhead{Replication of results}
Comprehensive descriptions of the simulations and computational methods are provided to ensure reproducibility. The thermal solver used in the heat-transfer examples will be made publicly available in the near future.

\bmhead{Author Contributions}
Derivations and simulations were carried out by Giuseppe Romano. Rodrigo Arrieta was responsible for preparing the numerical simulations of the photonic example. Steven G. Johnson supervised the research. Giuseppe Romano prepared the initial draft of the manuscript. All authors contributed to manuscript revision and approved the final version.

\bibliography{biblio}
\end{document}